\documentclass[aps,prd, twocolumn,groupedaddress,amsmath,amssymb,preprintnumbers,nofootinbib]{revtex4-2}

\usepackage[dvipdfmx]{graphicx}
\usepackage{dcolumn}
\usepackage{bm}
\usepackage{braket}
\usepackage{breqn}
\usepackage{xcolor}
\usepackage{hyperref}

\begin{document}

\preprint{KUNS-3079}

\title{Detecting Parity-Violating Gravitational Wave Backgrounds with \\ Pulsar Polarization Arrays}

\author{Qiuyue Liang}
\email{qiuyue.liang@ipmu.jp}
\affiliation{Kavli Institute for the Physics and Mathematics of the Universe (WPI)$,$ University of Tokyo$,$ Kashiwa 277-8583$,$ Japan}
\author{Kimihiro Nomura}
\email{k.nomura@tap.scphys.kyoto-u.ac.jp}
\affiliation{Department of Physics$,$ Kyoto University$,$ Kyoto 606-8502$,$ Japan} 
\author{Hidetoshi Omiya}
\email{omiya@tap.scphys.kyoto-u.ac.jp}
\affiliation{Department of Physics$,$ Kyoto University$,$ Kyoto 606-8502$,$ Japan} 

\date{\today}

\begin{abstract} 
Pulsar timing arrays probe isotropic stochastic gravitational wave (GW) backgrounds in the nanohertz band but are insensitive to its parity-violating component.
Motivated by recent progress in pulsar polarization arrays, we study the response of pulsar polarimetry to GWs and evaluate its potential to detect circular polarization in isotropic stochastic GW backgrounds, which characterizes parity violation.
Based on geometric optics, we derive the rotation of the polarization of electromagnetic waves induced by propagation through a GW background.
We show that the cross-correlation between pulsar timing and polarimetry signals isolates the circular polarization component from the GW intensity, 
sharing the same Hellings-Downs angular pattern.
With future facilities such as the SKA, timing-polarimetry correlations could reach sensitivities to the circular polarization of GWs comparable to those of the current astrometric methods.
\end{abstract} 


\maketitle

\section{Introduction}

Stochastic gravitational wave backgrounds (SGWBs) are prime targets of gravitational wave (GW) observations, as they carry rich information about the formation history of astrophysical compact objects and galaxies, as well as about the early universe (see, e.g., Refs.~\cite{Maggiore:1999vm, Christensen:2018iqi} for reviews). 
Recently, pulsar timing arrays (PTAs) have reported evidence for a common-spectrum process exhibiting Hellings--Downs correlations from a SGWB in the nanohertz band~\cite{NANOGrav:2023gor, EPTA:2023fyk, Reardon:2023gzh, Xu:2023wog}. 
At higher frequencies, ground-based interferometers (LIGO, Virgo, and KAGRA) have set upper limits on an isotropic background in the 10 Hz--kHz band~\cite{LIGOScientific:2025bgj}. 
In the future, space-based missions such as LISA~\cite{LISA:2017pwj}, Taiji~\cite{Hu:2017mde}, TianQin~\cite{Mei:2020lrl}, DECIGO~\cite{Kawamura:2020pcg}, and AEDGE~\cite{AEDGE:2019nxb} will open access to the milli- to decihertz range. 
Additional probes include astrometric searches~\cite{Book:2010pf}, photometric searches \cite{Wang:2020pmf}, asteroids ($\mu$Hz) \cite{Fedderke:2021kuy}, pulsar spin orbit (picoHz) \cite{Zheng:2025tcm}, cosmic microwave background (CMB) polarization measurements~\cite{LiteBIRD:2022cnt, CMB-S4:2020lpa}, and large-scale structure observations~\cite{Okumura:2024xnd,Mentasti:2024fgt}. 
Together, these efforts map the GW spectrum over a wide range of frequencies, with PTAs (and astrometry) uniquely sensitive to the nanohertz regime.

An intriguing open question is whether the SGWB violates parity, as such a violation would offer valuable insights into the physics of its origin.
A direct indicator of parity violation is the statistical circular polarization of GWs, since it quantifies the asymmetry between left- and right-handed modes.
Parity-violating GWs can arise in early-universe scenarios involving Chern--Simons gravity~\cite{Lue:1998mq, Alexander:2004us, Satoh:2007gn, Alexander:2009tp, Kanno:2023kdi}, Ho\v{r}ava--Lifshitz gravity~\cite{Takahashi:2009wc,Wang:2012fi,Zhu:2013fja,Qiao:2022mln, Li:2024fxy}, or parity violation in the matter sector~\cite{Sorbo:2011rz,Cook:2011hg,Dimastrogiovanni:2012ew, Adshead:2013nka, Dimastrogiovanni:2016fuu,Obata:2016tmo,Adshead:2018doq,Machado:2018nqk,Bastero-Gil:2022fme,Gerlach:2025fkr,Aoki:2025uwz}.
Primordial helical magnetic fields or turbulence may also generate circularly polarized GWs~\cite{Okano:2020uyr, Chiba:2025odu, Kahniashvili:2005qi}, and even astrophysical sources could produce a SGWB with non-negligible circular polarization~\cite{ValbusaDallArmi:2023ydl}.
Despite its potential significance for both fundamental physics and astrophysics, current tests provide no compelling evidence for parity violation in the SGWB. 
For prospects of future searches, see, e.g., Refs.~\cite{Seto:2008sr, Seto:2020zxw, Orlando:2020oko, Omiya:2021zif, Chen:2024fto, Mikura:2025vaj}.

In the nanohertz band, to which PTAs are sensitive, parity violation in the SGWB has so far been only weakly constrained.
The main difficulty is that PTAs are blind to the parity violation, i.e., the circular polarization, of an isotropic SGWB ~\cite{Kato:2015bye, Smith:2016jqs, Belgacem:2020nda,Sato-Polito:2021efu}.
This is essentially because PTAs measure a scalar quantity---the timing residual---for each pulsar, yielding a scalar map on the sky. 
For an isotropic background, scalar two-point correlations cannot distinguish left- from right-handed circular polarization, so timing-only cross-correlations are parity-even.
Parity-odd information can be accessed either by introducing a preferred direction (e.g., anisotropy in the background) or by constructing sky map of vector observables.
Both approaches have been discussed in the contexts of anisotropy searches with PTAs~\cite{Kato:2015bye, Belgacem:2020nda, Sato-Polito:2021efu,Cruz:2024esk}, in astrometric methods~\cite{Liang:2023pbj,Caliskan:2023cqm, Qin:2018yhy}, and in the shape of galaxies~\cite{Okumura:2024xnd}, but current sensitivities remain limited~\cite{Jaraba:2025hay}.
To facilitate the detection of parity violation in the SGWB, especially in the isotropic component, it is essential to employ additional observables.

In this paper, we propose a possible approach to detecting parity violation in the SGWB by combining PTAs with pulsar polarimetry.
Many millisecond pulsars are known to emit linearly polarized radiation~\cite{Lyne:1968kt, Yan:2011gh,Dai:2015awa,Guillemot:2023wfm}, which can remain stable over long timescales since the polarization angle is determined by the geometry of the pulsar beam~\cite{1969ApL.....3..225R,Ruderman:1975ju}. 
When metric perturbations are present in the form of GWs, photons propagating through this background experience a rotation of their polarization vector. 
This effect has been studied in the contexts of GW detection~\cite{Iacopini:1979ww, Cruise:2000za, Prasanna:2001xk, Faraoni:2007uh,Park:2020ugc,Shoom:2022oer, Garcia-Cely:2025mgu} and CMB polarization~\cite{Surpi:1997gh, Faraoni:2007uh, DiDio:2019rfy, Santana:2020bvu}, and is analogous to gravitational Faraday rotation in Kerr spacetime~\cite{Dehnen:1973xa, 1980ApJ...235..224C, Ishihara:1987dv}. 
The temporal variation in the polarization direction of each pulsar is typically monitored to measure the Faraday rotation induced by the galactic magnetic field, but it may also carry signatures of background GWs when observed over a long time baseline ($\sim 15$ years). 
In analogy with PTAs, one can ultimately assemble pulsar polarization arrays (PPAs), recording time series of polarizations for many pulsars and analyzing their correlations.
This concept has been explored in the context of axion-induced birefringence~\cite{Liu:2021zlt}, and has been used to place upper bounds on the axion-photon coupling in the mass range $10^{-24}$–$10^{-22}$ eV~\cite{EPTA:2024gxu, Li:2025xlr}. 
Since the rotation of polarization  angle is a pseudo-scalar rather than a scalar, it may open a pathway to accessing the parity-violating component of the SGWB.

Motivated by this, we study the effects of the SGWB on PPAs, as well as on PTAs. Especially, we focus on the cross-correlation between PPAs and PTAs, which has not been explored so far.
Intriguingly, we find that the cross-correlation between the pulsar timing redshift $z$ and the rotation of polarization  angle $\chi$ is sensitive to the circular polarization of the isotropic SGWB.
Moreover, the associated angular correlation pattern follows the Hellings--Downs curve. 
Consequently, the $z$-$\chi$ cross-correlation can serve as a new probe of parity violation in the SGWB.

The outline of this paper is as follows.
In Sec.~\ref{sec:2}, we derive expressions for the redshift and the rotation of polarization vectors of electromagnetic waves propagating through  background plane GWs.
In Sec.~\ref{sec:3}, we compute the cross-correlation between the redshift and the rotation of polarization  angle induced by the SGWB, and provide a forecast for the sensitivity to its circular polarization component.
We summarize our findings in Sec.~\ref{sec:4}.
Throughout this paper, we use the mostly-plus metric signature $(-,+,+,+)$ and adopt natural units with $c=1$.

\section{Redshift and polarization rotation of electromagnetic waves by plane GWs}
\label{sec:2}

\subsection{Setup of the problem}

In this section, we derive the redshift and the polarization rotation of electromagnetic waves induced by plane GWs.
We first present the setup of the problem.
We model the GW as a perturbation, $h_{ij}$, in Minkowski spacetime,
\begin{align}\label{eq:metric}
  g_{\mu\nu}dx^\mu dx^\nu = -dt^2 + [\delta_{ij} +h_{ij} (t, \bm{x}) ] dx^i dx^j~,
\end{align}
and adopt the transverse-traceless (TT) gauge, $h^i_{~i} = \partial_i h^i_{~j} = 0$.
Throughout this work, we restrict our analysis to linear order in $h_{ij}$.\footnote{We will focus only on linear perturbations throughout the paper; therefore, the summation convention between upper and lower spatial indices will not be specified.} We summarize the quantities to be derived in Table~\ref{tab:notation}.

\renewcommand{\arraystretch}{1.2}
\begin{table*}
\centering
\caption{Notation and definitions of the vectors used in this paper.
}
\label{tab:notation}
\begin{tabular}{ccc}\hline
Vector & Meaning  & Equation\\ \hline\vspace{1pt}
$\bar{k}^\mu$ & unperturbed wave vector & Eq.~\eqref{eq:unpertwv}\\ 
$\hat{n}^i$ & propagation direction of unperturbed photon & Eq.~\eqref{eq:unpertwv}\\ 
$\bar{\epsilon}^\mu$ & unperturbed polarization vector & Eq.~\eqref{eq:unpertpol}\\ 
$\bar{p}^i$ & reference direction for measurement of polarization& 
Eq.~\eqref{eq:barp}
\\ 
\hline \hline
$k^\mu(P)$ & perturbed wave vector at the pulsar& Eq.~\eqref{eq:kPulasar}\\ 
$k^\mu(E)$ & perturbed wave vector at the Earth& Eq.~\eqref{eq:kEarth}\\ 
$\epsilon^\mu$ & perturbed polarization vector & Eqs.~\eqref{eq:deltaeps0} and \eqref{eq:deltaepsi}\\
$e^\mu_{(\alpha)}(P)~ (\alpha = k, \parallel, \perp)$ & space-like reference frame at the pulsar & Eqs.~\eqref{eq:polbasiskP}--\eqref{eq:polbasis}
\\
$e^\mu_{(\alpha)}(E)~ (\alpha = k, \parallel, \perp)$ & space-like reference frame at the Earth & Eqs.~\eqref{eq:polparak}--\eqref{eq:polperp2}\\
\hline
\end{tabular}
\end{table*}

We consider the regime where the wavelength of the electromagnetic waves is much shorter than that of the GW, as is relevant for pulsar observations. 
In this limit, geometric optics applies to the electromagnetic waves: the photon wave vector $k^\mu$ obeys a null geodesic equation, and the polarization vector $\epsilon^\mu$ is perpendicular to the wave vector and is parallel transported along the light ray (see Appendix \ref{app:geometric-optics}), 
\begin{align}
\label{eq:geodlight}
  k^\mu\nabla_\mu k^\nu=0~, \qquad g_{\mu\nu} k^\mu k^\nu=0~, \\
\label{eq:polprop}
  k^\mu\nabla_\mu \epsilon^\nu=0~, \qquad g_{\mu\nu} k^\mu \epsilon^\nu=0~,
\end{align}
where $\nabla_\mu$ denotes the covariant derivative associated with the metric \eqref{eq:metric}.
The polarization vector $\epsilon^\mu$ is complex in general and is defined with the normalization condition,
\begin{align}
    g_{\mu\nu} \epsilon^\mu \epsilon^{\nu *} = 1 \,,
\end{align}
where the asterisk (*) denotes the complex conjugate.
Note that there is a gauge freedom $\epsilon^\mu\to\epsilon^\mu+\alpha k^\mu$ with constant $\alpha$.

Let us express the solutions for $k^\mu$ and $\epsilon^\mu$ as
\begin{align}
\label{eq:wavevector}
    k^\mu &= \bar{k}^\mu + \delta k^\mu \,,\\
    \epsilon^\mu &= \bar{\epsilon}^\mu + \delta \epsilon^\mu \,,
\end{align}
where $\bar{k}^\mu$ and $\bar{\epsilon}^\mu$ denote the solutions in Minkowski spacetime, and $\delta k^\mu$ and $\delta \epsilon^\mu$ represent the perturbations induced by the GW. 
In this section, we will find $\delta k^\mu$ and $\delta \epsilon^\mu$ to linear order in $h_{ij}$.

We study a setup in which a pulsar emits electromagnetic waves that are observed on Earth.
For simplicity, we treat both the observer and the pulsar as moving along timelike geodesics.
In the TT gauge, the worldlines of fixed spatial coordinates satisfy the geodesic equation. 
Therefore, even in the presence of the GW, we can assume that the observer (Earth) is located at the spatial origin, $\bm{x}_{E} = \bm{0}$, and the pulsar remains at a fixed spatial position, $\bm{x}_P = L \,\hat{\bm{n}}$, where $L$ is the distance between the Earth and the pulsar, and $\hat{\bm{n}}$ is the unit vector pointing from the Earth to the pulsar, satisfying $\delta_{ij} \hat{n}^i \hat{n}^j = 1$.
In this setup, the four-velocities of the observer (Earth) $u_E^\mu$ and the pulsar $u_P^\mu$ are both given by $u_E^\mu = u_P^\mu = (1, \bm{0})$.

In the unperturbed Minkowski spacetime, the photon propagates in the direction $-\hat{\bm{n}}$, with a constant null wave vector,
\begin{align}\label{eq:unpertwv}
    \bar{k}^\mu = \omega_P (1, -\hat{\bm{n}})~,
\end{align}
where $\omega_P$ denotes the angular frequency of the emitted light. 
The unperturbed trajectory of the photon, $\bar{x}^\mu$, is obtained by integrating $d\bar{x}^\mu / d\lambda = \bar{k}^\mu$ with respect to the affine parameter $\lambda$, yielding
\begin{align}
    \bar{x}^\mu (\lambda) 
    &= (t_{E} + \omega_P (\lambda - \lambda_E ) , - \omega_P (\lambda - \lambda_E ) \hat{\bm{n}}) \,.
\end{align}
Here, the light ray departs from the pulsar at $\lambda=\lambda_P = 0$ and reaches the Earth at $\lambda = \lambda_E \equiv L/\omega_P$.\footnote{Our convention of the affine parameter differs from that of Book and Flanagan \cite{Book:2010pf}, in which $\lambda = 0$ is set at Earth and $\lambda = -L/\omega_P$ at the pulsar.} 
We take the photons to be observed on Earth at $t = t_{E}$, so that the emission time at the pulsar is $t_{P} = t_{E} - L$.

\subsection{Perturbed light ray and redshift}

Let us first solve the geodesic equation~\eqref{eq:geodlight} to linear order in $h_{ij}$. 
The linearized geodesic equation reads
\begin{align}\label{eq:pertk}
  \frac{d\,\delta k^\mu}{d\lambda}=-{\Gamma}^\mu_{~\nu\rho}\,\bar{k}^\nu\bar{k}^\rho ,
\end{align}
where ${\Gamma}^\mu_{~\nu\rho}$ denotes the Christoffel symbol.
For the metric \eqref{eq:metric} including the GW, the components of Eq.~\eqref{eq:pertk} take the following form,
\begin{align}
    \frac{d \, \delta k^0}{d \lambda} &= - \frac{\omega_P^2}{2} \, \partial_0 {h}_{ij}\,  \hat{n}^i \hat{n}^j \,, \\
    \frac{d \, \delta k^i}{d \lambda} &= 
    \omega_P^2 \, \partial_0 {h}_{ij} \, \hat{n}^j - \frac{\omega_P^2}{2} ( \partial_k h_{ij} + \partial_j h_{ik} - \partial_i h_{jk}) \,\hat{n}^j \hat{n}^k \,.  
\end{align}
Integrating these equations yields
\begin{align}
  \delta k^0(\lambda) &= -\frac{\omega_P^2}{2}\,I_{ij}(\lambda)\,\hat{n}^i\hat{n}^j + C^0 ~, 
  \label{eq:pertksol0}\\
  \delta k^i(\lambda) &= \omega_P \, \Delta h_{ij}(\lambda) \,\hat{n}^j
    + \frac{\omega_P^2}{2}J_{jki}(\lambda) \,\hat{n}^j\hat{n}^k + C^i ~, 
    \label{eq:pertksol}
\end{align}
with integration constants $C^0$ and $C^i$, and
\begin{align}
  \Delta h_{ij}(\lambda) &\equiv h_{ij}(\lambda)-h_{ij}(0)~,\\
  I_{ij}(\lambda) &\equiv \int_0^\lambda d\lambda'\,\partial_0 h_{ij}(\lambda')~,\\
 J_{jki} (\lambda) &\equiv \int_0^\lambda d\lambda'\,\partial_i h_{jk}(\lambda')~.
\end{align}
Here, $h_{ij}(\lambda) \equiv h_{ij}(\bar{x}^\mu(\lambda))$.
The integrals can be evaluated along the unperturbed trajectory $\bar{x}^\mu(\lambda)$, since we consider only to linear order in $h_{ij}$.
Moreover, in Eq.~\eqref{eq:pertksol}, we used the  relation $d/d\lambda = \omega_P (\partial_0 - \hat{n}^i \partial_i)$, which holds along the unperturbed trajectory.
The integration constants $C^0$ and $C^i$ will be determined later.

Integrating $d x^\mu / d\lambda = k^\mu$ with $k^\mu = \bar{k}^\mu + \delta k^\mu$, we obtain the perturbed trajectory, $x^\mu(\lambda)$, to linear order in $h_{ij}$ as
\begin{align}
  x^\mu(\lambda)=\bar{x}^\mu(\lambda)+\int_0^\lambda d\lambda'\,\delta k^\mu(\lambda')+ X^\mu ~,
\end{align}
where $X^\mu$ are integration constants.

The eight integration constants $C^0, C^i$ and $X^\mu$ are determined by the following four conditions \cite{Book:2010pf}: 1) the light ray intersects the worldline of the pulsar ($\bm{x}_P = L \,\hat{\bm{n}}$) at $\lambda_P = 0$; 2) the emitted angular frequency in the pulsar's local frame is $\omega_P$;
3) the light ray reaches the Earth at $x_E^\mu = (t_{E}, \bm{0})$; and 4) the wave vector $k^\mu$ is null. These conditions lead to $X^i = 0$, $C^0=0$, and
\begin{align}
  C^i &= 
  - \frac{\omega_P}{\lambda_E} \,\hat{n}^j \int_{0}^{\lambda_E} d\lambda' \, \Delta h_{ij}(\lambda') 
  \notag \\
  &\quad 
    + \frac{\omega_P}{2\lambda_E} \hat{n}^i \hat{n}^j \hat{n}^k \int_{0}^{\lambda_E} d\lambda' \, [ h_{jk}(\lambda') + \omega_P I_{jk}(\lambda') ]
  \notag \\
  &\quad 
    - \frac{\omega_P^2}{2\lambda_E} \hat{n}^j \hat{n}^k \int_{0}^{\lambda_E} d\lambda' \,J_{jki}(\lambda') ~,
  \\
  X^{0} &= 
  - \frac{\omega_P}{2} \hat{n}^i \hat{n}^j \int_0^{\lambda_E} d\lambda' \, h_{ij}(\lambda') ~.
\end{align}
Note that we define the origin of the affine parameter, $\lambda = 0$, to correspond to the emission at the pulsar even in the presence of the GW. In this convention, the affine parameter value at the Earth is shifted from $\lambda_E = L/\omega_P$ to $\tilde{\lambda}_E$ due to the GW, which is given by
\begin{align}
    \tilde{\lambda}_E = \lambda_E + \frac{1}{2} \hat{n}^i \hat{n}^j \int_0^{\lambda_E} d\lambda' \, [ h_{ij}(\lambda') + \omega_P I_{ij}(\lambda') ]~.
\end{align}

Evaluating Eqs.~\eqref{eq:pertksol0} and \eqref{eq:pertksol} at $\lambda = 0$ and $\lambda = \tilde{\lambda}_E$, we obtain the wave vectors, Eq.\eqref{eq:wavevector}, at the pulsar ($P$) and at the Earth ($E$), to linear order in $h_{ij}$, as
\begin{align}
  k^\mu(P) &= \bar{k}^\mu + (0, C^i ) ~,\label{eq:kPulasar}\\
  k^\mu(E) &= \bar{k}^\mu 
  + \bigg( - \frac{\omega_P^2}{2}I_{ij}(\lambda_E) \, \hat{n}^i\hat{n}^j ~, \cr
  &
    \omega_P \, \Delta h_{ij}(\lambda_E) \,\hat{n}^j
    + \frac{\omega_P^2}{2}J_{jki}(\lambda_E) \,\hat{n}^j\hat{n}^k + C^i
    \bigg)\ , 
    \label{eq:kEarth}
\end{align}
respectively.

The angular frequency of the photons observed on Earth is given by $\omega_E = - g_{\mu\nu} (E) \,u_{E}^\mu k^\nu(E)$, where $g_{\mu\nu}(E)$ denotes the metric evaluated at the observation point, and $u_{E}^\mu=(1,\bm{0})$. 
To linear order in $h_{ij}$, it reads 
\begin{align}
    \omega_E = \omega_P - \frac{\omega_P^2}{2}  I_{ij}(\lambda_E) \, \hat{n}^i \hat{n}^j ~.
    \label{eq:omegaE}
\end{align}
The fractional frequency shift, i.e., the redshift $z$, is then
\begin{align}
  z \equiv \frac{\omega_P-\omega_E}{\omega_P}
  = \frac{\omega_P}{2} \hat{n}^i\hat{n}^j  \int_{0}^{\lambda_E} d\lambda \, \partial_0 h_{ij}(\lambda) \,, 
\end{align}
which reproduces the well-known result~\cite{Book:2010pf}. 
Changing the integration variable from $\lambda$ to $t = \bar{t}(\lambda)$, we obtain 
\begin{align}
    z(t_E) = \frac{1}{2} \hat{n}^i\hat{n}^j  \int_{t_E - L}^{t_E} dt \, \partial_0 h_{ij}(t, \bm{x}) \Big|_{\bm{x} = \bar{\bm{x}}(t)}\,,
    \label{eq:z_time}
\end{align}
where $t_{E}$ is the observation time.
In the integrand, the unperturbed trajectory $\bar{\bm{x}}(t) \equiv (t_{E} - t) \hat{\bm{n}}$ is substituted after evaluating the partial derivative with respect to $t$.
This redshift is the observable in PTAs as deviations in the pulse periods.

\subsection{Polarization rotation}

We now turn to the evolution of the polarization vector $\epsilon^\mu$ along the light ray.
In the unperturbed Minkowski spacetime, the transport equation is simply given by $d \bar{\epsilon}^\mu / d\lambda = 0$, whose solution is $\bar{\epsilon}^\mu = \text{const}$.
Exploiting the gauge freedom, we can further choose
\begin{align}\label{eq:unpertpol}
    \bar{\epsilon}^0 = 0 \,, \qquad 
    \bar{\epsilon}^i = \text{const}~.
\end{align}

Moreover, by assuming linearly polarized light (we will show later that this is justified), we regard $\bar{\epsilon}^i$ as real.
The vector $\bar{\epsilon}^i$ satisfies the normalization condition in flat space, 
\begin{align}
    \delta_{ij} \bar{\epsilon}^i \bar{\epsilon}^j =1 ~.
\end{align}

Let us study the perturbation $\delta \epsilon^\mu$ induced by the GW. To linear order in $h_{ij}$, the parallel transport equation \eqref{eq:polprop} reduces to
\begin{align}
  \frac{d \, \delta \epsilon^\mu}{d\lambda}= - {\Gamma}^\mu_{~\nu\rho}\,\bar{k}^\nu \bar{\epsilon}^\rho ~.
\end{align}
Using the metric \eqref{eq:metric}, the components of this equation can be written as 
\begin{align}
    \frac{d \, \delta \epsilon^0}{d \lambda} &= \frac{\omega_P}{2} \, \partial_0 h_{ij} \, \hat{n}^i \bar{\epsilon}^j \,,
    \\
    \frac{d \, \delta \epsilon^i}{d\lambda} &= - \frac{\omega_P}{2} \, \partial_0 h_{ij} \, \bar{\epsilon}^j  + 
    \frac{\omega_P}{2} ( \partial_k h_{ij} + \partial_j h_{ik} - \partial_i h_{jk}) \hat{n}^j \bar{\epsilon}^k \,.
\end{align}
Integrating these equations along the light ray gives
\begin{align}
    \delta \epsilon^0(\lambda) &= \frac{\omega_P}{2} I_{ij}(\lambda) \, \hat{n}^i  \bar{\epsilon}^j + D^0 ~,
    \label{eq:deltaeps0}
    \\
    \delta \epsilon^i (\lambda)
    &= - \frac{1}{2} \Delta h_{ij}(\lambda) \, \bar{\epsilon}^j 
    + \frac{\omega_P}{2} J_{j[ik]}(\lambda) \, \hat{n}^j \bar{\epsilon}^k + D^i ~,
    \label{eq:deltaepsi}
\end{align}
where $D^0$ and $D^i$ are integration constants determined later, and the brackets denote anti-symmetrization on the last two indices, $J_{j[ik]} = J_{jik} - J_{jki}$.

To define the polarization angle at the pulsar and at the Earth, it is necessary to introduce a reference frame at each location.
First, the reference frame at the pulsar can be constructed as follows.
At the pulsar, we introduce a space-like reference vector $p_P^\mu$. 
We require that $p_P^\mu$ is normalized as $g_{\mu\nu} p_P^\mu p_P^\nu = 1$, orthogonal to the pulsar's four-velocity, $g_{\mu\nu} u_P^\mu p_P^\nu = 0$, and parallel transported along the worldline of the pulsar, $u_P^\mu \nabla_\mu p_P^\nu=0$.
Let us write $p_P^\mu = \bar{p}^\mu + \delta p_P^\mu$, where $\bar{p}^\mu$ denotes the reference vector in Minkowski spacetime and $\delta p_P^\mu$ is the perturbation due to the GW.
Obviously, the unperturbed vector $\bar{p}^\mu$ is constant.
We can choose $\bar{p}^\mu$ to be orthogonal to both the pulsar's four-velocity and the light ray in Minkowski spacetime.
Therefore, $\bar{p}^\mu$ is expressed as $\bar{p}^\mu = (0, \bar{\bm{p}})$ with
\begin{align}
    \delta_{ij} \bar{p}^i \bar{p}^j = 1~, 
    \qquad 
    \delta_{ij} \hat{n}^i \bar{p}^j = 0~.
    \label{eq:barp}
\end{align}
Solving the parallel transport equation $u_P^\mu \nabla_\mu p_P^\nu=0$ with $u_P^\mu = (1, \bm{0})$, we obtain, to linear order in $h_{ij}$, the reference vector at the pulsar as
\begin{align}
  p_P^\mu &= \bigg( 0,~\bar{p}^i-\tfrac{1}{2}h_{ij}(P) \,\bar{p}^j  \bigg)~, 
\end{align}
where $h_{ij}(P)$ denotes the metric perturbation at the pulsar's position, and integration constants have been omitted since they are irrelevant to the polarization rotation.

Using the reference vector $p_P^\mu$, we can introduce a tetrad associated with the pulsar by the Gram-Schmidt orthogonalization as follows~\cite{2012ApJ...752..123G}. 
We identify the four-velocity as the timelike basis vector; $e_{(t)}^\mu(P) = u_P^\mu$.
Using the photon's wave vector at the pulsar, $k^\mu(P)$, a space-like basis vector orthogonal to $e_{(t)}^\mu(P)$ is given by 
\begin{align}
    e^\mu_{(k)}(P) &= \frac{k^\mu(P)}{\omega_P}-u_P^\mu ~.
\end{align}
Notice that $\omega_P = - u_P^\mu k_\mu(P)$.
The other two orthonormal basis vectors can be obtained using $p_P^\mu$ as 
\begin{align}
    e_{(\parallel)}^\mu(P) &= \frac{p_P^\mu - ( p_P^\nu k_\nu(P) / \omega_P ) \, e_{(k)}^\mu(P)}{[ 1 - ( p_P^\lambda k_\lambda(P) / \omega_P )^2 ]^{1/2}}~,
    \\
    e_{(\perp)}^\mu(P) &= {\varepsilon^\mu}_{\nu\rho\sigma} u_P^\nu \, e_{(k)}^\rho(P) \, e_{(\parallel)}^\sigma(P) 
    \notag \\
    &= \frac{{\varepsilon^\mu}_{\nu\rho\sigma} u_P^\nu k^\rho(P) \, p_P^\sigma}{[ \omega_P^2 - (p_P^\lambda k_\lambda(P))^2 ]^{1/2}}~.
\end{align}
Here, $\varepsilon_{\mu\nu\rho\sigma}$ is the antisymmetric tensor normalized as $\varepsilon_{0123} = +\sqrt{|\det[g_{\mu\nu}]|}$.
These basis vectors satisfy $g_{\mu\nu}(P) \, e_{(\alpha)}^\mu(P) \, e_{(\beta)}^\nu(P) = \eta_{\alpha\beta}$ with $(\alpha, \beta) = (t, k, \parallel, \perp)$, where $\eta_{\alpha\beta}$ is the Minkowski metric.
The basis vectors are expanded to linear order in $h_{ij}$ as 
\begin{align}
    e_{(k)}^\mu (P)&= \bigg( 0, - \hat{n}^i + \frac{\delta k^i(P)}{\omega_P} \bigg)  ~,
    \\
    e_{(\parallel)}^\mu (P)&= \bigg( 
    0, \bar{p}^i - \frac{1}{2} h_{ij}(P) \, \bar{p}^j 
    \notag \\
    &\qquad 
    + \hat{n}^i
    \bigg[ \frac{\delta k^j(P) \, \bar{p}_j \, }{\omega_P} - \frac{1}{2} h_{jk}(P) \, \bar{p}^j \hat{n}^k \bigg]
    \bigg)~,
    \label{eq:eparaP}
    \\
    e_{(\perp)}^\mu (P)
    &= \bigg( 0, \bar{\varepsilon}_{ijk} \hat{n}^j \bar{p}^k 
    - h_{ij}(P) \, \bar{\varepsilon}_{jkl} \,\hat{n}^k \bar{p}^l
    \notag \\
    &\qquad - 
    \bar{\varepsilon}_{ijk} \bigg[ \frac{\delta k^j(P) \, \bar{p}^k}{\omega_P} + \frac{1}{2} \hat{n}^j h_{kl}(P) \, \bar{p}^l \bigg]
    \bigg) ~,
\end{align}
where $\bar{\varepsilon}_{ijk}$ is the three-dimensional antisymmetric tensor in flat space normalized as $\bar{\varepsilon}_{123} = +1$. 
Note that $e_{(\parallel)}^\mu$ reduces to $\bar{p}^\mu = (0, \bar{\bm{p}})$ in the absence of the GW.

Similarly, at the Earth, we introduce a spatial reference vector $p_E^\mu$ that satisfies $g_{\mu\nu} p_E^\mu p_E^\nu = 1$, $g_{\mu\nu} u_E^\mu p_E^\nu = 0$, and $u_E^\mu \nabla_\mu p_E^\nu=0$, where $u_E^\mu$ is the four-velocity of the observer (Earth).
We write $p_E^\mu = \bar{p}^\mu + \delta p_E^\mu$, where $\bar{p}^\mu$ is identical to that in Eq.~\eqref{eq:barp}, and the perturbation $\delta p_E^\mu$ is derived from the parallel transport equation.
To linear order in $h_{ij}$, we obtain 
\begin{align}
      p_E^\mu &= \bigg( 0,\bar{p}^i-\frac{1}{2}h_{ij}(E) \, \bar{p}^j \bigg)~,
      \label{eq:refdirE}
\end{align}
where $h_{ij}(E)$ denotes the metric perturbation at the Earth's position, and, again, we have omitted the integration constants that are irrelevant to the polarization rotation.
We can construct a tetrad associated with the Earth as $e_{(t)}^\mu(E) = u_E^\mu$, and 
\begin{align}
    e^\mu_{(k)}(E) &= \frac{k^\mu(E)}{\omega_E}-u_E^\mu~, \label{eq:polbasiskP}\\
    e^\mu_{(\parallel)}(E) &= \frac{p_E^\mu -(p_E^\nu k_\nu(E) / \omega_E) \, e^\mu_{(k)}(E)}{ [ 1 - ( p_E^\lambda k_\lambda(E) / \omega_E)^2 ]^{1/2}}~, \\
  e^\mu_{(\perp)}(E) &= {\varepsilon^{\mu}}_{\nu\rho\sigma} u_E^\nu \, e^\rho_{(k)}(E) \, e^\sigma_{(\parallel)}(E) 
  \notag \\
  &= \frac{{\varepsilon^{\mu}}_{\nu\rho\sigma} u_E^\nu \, k^\rho(E) \, p_E^\sigma}{ [ \omega_E^2 - (p_E^\lambda k_\lambda(E))^2 ]^{1/2} }
  ~,
  \label{eq:polbasis}
\end{align}
where $k^\mu(E)$ is the photon's wave vector at the Earth \eqref{eq:kEarth}, and $\omega_E = - u_E^\mu k_\mu(E)$ is the observed angular frequency \eqref{eq:omegaE}.
Expanding to linear order in $h_{ij}$ yields
\begin{align}
    e_{(k)}^\mu(E) &= \bigg( 0, - \frac{\omega_P}{\omega_E} \hat{n}^i + \frac{\delta k^i (E)}{\omega_P} \bigg) ~,\label{eq:polparak}
    \\
    e_{(\parallel)}^\mu(E) &= \bigg( 0 , \bar{p}^i - \frac{1}{2} h_{ij}(E) \, \bar{p}^j 
    \notag \\
    &\qquad + \hat{n}^i 
    \bigg[ \frac{\delta k^j(E) \, \bar{p}_j}{\omega_P} - \frac{1}{2} h_{jk}(E) \, \bar{p}^j \hat{n}^k \bigg] 
    \bigg) ~, 
    \label{eq:polpara2}\\
    e_{(\perp)}^\mu (E)
    &= \bigg( 0, \frac{\omega_P}{\omega_E} \bar{\varepsilon}_{ijk} \hat{n}^j \bar{p}^k 
    - h_{ij}(E) \, \bar{\varepsilon}_{jkl} \, \hat{n}^k \bar{p}^l
    \notag \\
    &\qquad - 
    \bar{\varepsilon}_{ijk} \bigg[ \frac{\delta k^j(E) \, \bar{p}^k}{\omega_P} + \frac{1}{2} \hat{n}^j h_{kl}(E) \, \bar{p}^l \bigg]
    \bigg) ~.
    \label{eq:polperp2}
\end{align}

By construction, the basis vectors $e_{(\parallel)}^\mu(P)$ and $e_{(\perp)}^\mu(P)$ are orthogonal to the photon's wave vector at the pulsar. 
Accordingly, the polarization vector at the pulsar, $\epsilon^\mu(P)$, can be expressed as a linear combination of these two basis vectors.
The same property also holds at the Earth.

In what follows, we will assume that the pulsar emits linearly polarized light whose polarization is parallel to the basis vector  $e_{(\parallel)}^\mu(P)$:
\begin{align}
    \epsilon^\mu(P) = e^\mu_{(\parallel)}(P)~.
    \label{eq:polcond}
\end{align}
In general, the initial polarization may have a nonzero angle with respect to $e_{(\parallel)}^\mu(P)$, and circular polarization may also be present. However, our results will show that the initial polarization angle is irrelevant to the polarization rotation, and circular polarization is not affected by the GWB.
For the unperturbed component, Eq.~\eqref{eq:polcond} indicates that 
\begin{align}
    \bar{\epsilon}^i = \bar{p}^i ~.
\end{align}
Meanwhile, by substituting Eqs.~\eqref{eq:deltaeps0} and \eqref{eq:deltaepsi} evaluated at the pulsar ($\lambda = 0$) and Eq.~\eqref{eq:eparaP} into Eq.~\eqref{eq:polcond}, the integration constants $D^0$ and $D^i$ are determined as 
\begin{align}
    D^0 &= 0 ~, \\
    D^i &= - \frac{1}{2} h_{ij}(P) \, \bar{\epsilon}^j 
    + \hat{n}^i
    \bigg[ \frac{\delta k^j(P) \, \bar{\epsilon}_j \, }{\omega_P} - \frac{1}{2} h_{jk}(P) \, \bar{\epsilon}^j \hat{n}^k \bigg]~.
\end{align}
Using these results, the polarization vector at the Earth can be obtained as 
\begin{align}
    \epsilon^0(E) &= \frac{\omega_P}{2} I_{ij}(\lambda_E) \, \hat{n}^i \bar{\epsilon}^j ~,
    \\
    \epsilon^i(E) &= \bar{\epsilon}^i - \frac{1}{2} h_{ij}(E) \, \bar{\epsilon}^j 
    + \frac{\omega_P}{2} J_{j[ik]} (\lambda_E) \, \hat{n}^j \bar{\epsilon}^k
    \notag \\
    &\quad 
    + \hat{n}^i
    \bigg[ \frac{\delta k^j(P) \, \bar{\epsilon}_j \, }{\omega_P} - \frac{1}{2} h_{jk}(P) \, \bar{\epsilon}^j \hat{n}^k \bigg] ~.
    \label{eq:polvecE}
\end{align}
One can directly verify that this $\epsilon^\mu(E)$ is indeed orthogonal to $k^\mu(E)$ given in Eq.~\eqref{eq:kEarth}. 
Moreover, we see that when the unperturbed polarization vector $\bar{\epsilon}^i$ is real, the perturbed polarization vector also remains real. Consequently, no circular polarization of the photons is generated.

We now relate the above discussion to the rotation of the linear polarization angle in pulsar array systems. In geometric-optics limit, the photon field $A^\mu$ is expressed as $A^\mu = a\, \epsilon^\mu \, e^{i\theta}$ with the scalar amplitude $a$ and the phase $\theta$ (see Appendix \ref{app:geometric-optics}).
The wave vector is then defined as the gradient of the phase, $k^\mu \equiv \nabla^\mu \theta$. 
At leading order in geometric-optics, $|\nabla^\mu a| \ll |\nabla^\mu \theta| $, the field strength tensor $F^{\mu\nu} \equiv \nabla^\mu A^\nu - \nabla^\nu A^\mu$ can be expressed as
\begin{align}
    F^{\mu\nu} &= i\, a \, ( k^\mu \epsilon^\nu - k^\nu \epsilon^\mu ) \, e^{i\theta}\,.
    \label{eq:Fmunu}
\end{align}
Then, the electric field vector in the observer’s inertial frame, $E^\mu$, is defined using the observer's four-velocity $u_E^\mu$ as
\begin{align}
    E^\mu \equiv u_E^\nu \, {F^\mu}_{\nu}(E) ~,
    \label{eq:defE}
\end{align}
where $F^{\mu\nu}(E)$ is the field strength tensor evaluated at the observation point (Earth). 
From $u_E^\mu = (1, \bm{0})$ and Eq.~\eqref{eq:Fmunu}, we find $E^0 = 0$ and, to linear order in $h_{ij}$,
\begin{align}
    E^j = i \, \omega_P \, a \, e^{i\theta}
    \bigg( \bar{\epsilon}^j + \frac{\delta k^0(E)}{\omega_P} \bar{\epsilon}^j + \delta \epsilon^j(E) + \hat{n}^j \, \delta \epsilon^0(E) \bigg) ~.
    \label{eq:obsE}
\end{align}

The electric field vector measured by the observer lies in the two-dimensional plane spanned by $e_{(\parallel)}^\mu(E)$ and $e_{(\perp)}^\mu(E)$.
Thus, we can write
\begin{align}
    E^\mu = E_{(\parallel)} \, e_{(\parallel)}^\mu(E) + E_{(\perp)} \, e_{(\perp)}^\mu(E) ~,
\end{align}
where
\begin{align}
    E_{(\parallel)} &= g_{\mu\nu}(E) \, E^\mu \, e_{(\parallel)}^\nu (E) ~,\\
    E_{(\perp)} &= g_{\mu\nu}(E) \, E^\mu \, e_{(\perp)}^\nu (E) ~.
\end{align}
Using Eqs.~\eqref{eq:polpara2}, \eqref{eq:polperp2} and \eqref{eq:obsE}, and noting that $\delta_{ij} \hat{n}^i \bar{\epsilon}^j = 0$, these components are calculated as 
\begin{align}
    E_{(\parallel)} &= 
    i \, \omega_P \, a \, e^{i\theta}
    \bigg( 1 + \frac{1}{2} h_{ij}(E) \, \bar{\epsilon}^i \bar{\epsilon}^j + \frac{\delta k^0(E)}{\omega_P} + \bar{\epsilon}^i \, \delta \epsilon_i(E) \bigg) 
    \notag \\
    &= i \, \omega_P \, a \, e^{i\theta}
    \bigg( 1 + \frac{\delta k^0(E)}{\omega_P} \bigg) 
    ~,
    \label{eq:Epara}\\
    E_{(\perp)} &= i \, \omega_P \, a \, e^{i\theta}
    \bigg( \bar{\varepsilon}_{ijk} \, \delta \epsilon^i(E) \, \hat{n}^j \bar{\epsilon}^k
    - \frac{1}{2} \bar{\varepsilon}_{ijk} \bar{\epsilon}^i \hat{n}^j h_{kl}(E)\, \bar{\epsilon}^l 
    \bigg)
    \notag \\
    &= i \, \omega_P \, a \, e^{i\theta} 
    \bigg( \frac{\omega_P}{2} J_{j[ik]} (\lambda_E) \, \hat{m}^i \hat{n}^j \bar{\epsilon}^k \bigg)~.
    \label{eq:Eperp}
\end{align}
In the second line of each equation, we have used Eq.~\eqref{eq:polvecE}. We have also introduced a unit vector in three-dimensional Euclidean space, $\hat{m}^i$, which is orthogonal to both $\hat{n}^i$ and $\bar{\epsilon}^i$:
\begin{align}
    \hat{m}^i \equiv \bar{\varepsilon}_{ijk} \hat{n}^j \bar{\epsilon}^k\,.
\end{align}

In terms of $E_{(\parallel)}$ and $E_{(\perp)}$, the photon's polarization angle $\chi$ measured from the $e_{(\parallel)}^i(E)$-axis is given by (see, e.g., Section 2.4 of Ref.~\cite{1979rpa..book.....R})
\begin{align}
    \tan 2 \chi = \frac{2 \, \text{Re} \big[ E_{(\parallel)} E_{(\perp)}^* \big] }{|E_{(\parallel)}|^2 - |E_{(\perp)}|^2} \,,
    \label{eq:defchi}
\end{align}
where $\text{Re}[\cdots]$ stands for the real part.\footnote{
The definition of the polarization angle $\chi$ in Eq.~\eqref{eq:defchi} is general and applies also to elliptically polarized light.
For linearly polarized light, Eq.~\eqref{eq:defchi} reduces to $\tan \chi = |E_{(\perp)}|/|E_{(\parallel)}|$.
}
Substituting Eqs.~\eqref{eq:Epara} and \eqref{eq:Eperp}, and assuming that $\chi$ is small, we obtain the rotation of polarization angle induced by the GW as
\begin{align}
  \chi = \frac{\omega_P}{2} \bar{\epsilon}^i 
  \hat{m}^j \hat{n}^k  \int_0^{\lambda_E}  d\lambda\,
  \partial_{[i}h_{j]k}(\lambda)~.
  \label{eq:polangle}
\end{align}
As in the case of the redshift \eqref{eq:z_time}, we can change the integration variable from $\lambda$ to $t = \bar{t}(\lambda)$, yielding 
\begin{align}
      \chi(t_{E})=\frac{1}{2}
      \bar{\epsilon}^i \hat{m}^j \hat{n}^k \int_{t_{E}-L}^{ t_{E} } dt\,
  \partial_{[i}h_{j]k} (t,\bm{x}) \Big|_{\bm{x} = \bar{\bm{x}}(t)}~.
  \label{eq:chi_time}
\end{align}
Since $h_{ij}$ depends on time, $\chi(t_{E})$ is observed as a polarization angle varying with the observation time $t_{E}$.

Notice that the second term in Eq.~\eqref{eq:polvecE} is completely canceled by the rotation of the basis vector induced by the GW, which is the $\mathcal{O}(h_{ij})$ term in Eq.~\eqref{eq:refdirE}.
It should also be noted that, although the unperturbed polarization vector $\bar{\epsilon}^i$ appears in Eq.~\eqref{eq:polangle}, the result is in fact independent of its orientation.
Indeed, by introducing any orthonormal basis vectors $(\bar{e}_{(1)}^i, \bar{e}_{(2)}^i)$ 
in the two-dimensional plane orthogonal to $\hat{n}^i$ as linear combinations of $\bar{\epsilon}^i$ and $\hat{m}^i$, the rotation of polarization angle $\chi$ can alternatively be written as
\begin{align}
    \chi = \frac{\omega_P}{2} \bar{\varepsilon}^{ab} \bar{e}_{(a)}^i \bar{e}_{(b)}^j \hat{n}^k 
    \int_0^{\lambda_E} d\lambda \, \partial_i h_{jk}(\lambda)~,
\end{align}
where $\bar\varepsilon^{ab}$ is the antisymmetric tensor in the two-dimensional tetrad plane with $\bar\varepsilon^{12} = +1$. 
It is clear that $\chi$ transforms as a pseudo-scalar under parity. Moreover, this expression remains invariant under a rotation of the basis vectors in that plane.
Thus, the initial polarization angle is irrelevant, which allows us to choose $\bar{p}^i$ to be aligned with the unperturbed polarization direction.

\subsection{Response functions}

We now show how the observables---the redshift and the rotation of polarization angle---respond to GWs. 
We express the GWs as a superposition of plane waves,
\begin{align}
  h_{ij}(t,\bm{x})&=\sum_{P=+,\times}\int_{-\infty}^{\infty} df \int d^2\hat{\bm{\Omega}} \,
  \tilde{h}_P(f,\hat{\bm{\Omega}}) 
  \notag \\
  &\quad \times e^P_{ij}(\hat{\bm{\Omega}})\,e^{- 2\pi i f(t-\hat{\bm{\Omega}}\cdot\bm{x})}~,
  \label{eq:pwave}
\end{align}
where $f$ is the frequency, $\hat{\bm{\Omega}}$ is the unit vector along the propagation direction of each plane wave, and $P (= +, \times)$ labels the GW polarizations.
For each propagation direction $\hat{\bm{\Omega}}$, the polarization tensors $e^P_{ij}(\hat{\bm{\Omega}})$ are constructed as
\begin{align}\label{eq:polGW}
    e^+_{ij}&=\hat{u}_i\hat{u}_j-\hat{v}_i\hat{v}_j~,& e^\times_{ij} &=\hat{u}_i\hat{v}_j+\hat{v}_i\hat{u}_j~,
\end{align}
where $\hat{\bm{u}}$ and $\hat{\bm{v}}$, together with $\hat{\bm{\Omega}}$, form an orthonormal triad. 
They are parametrized as
\begin{align}
  \hat{\bm{\Omega}}&=(\sin\theta\cos\phi, \sin\theta\sin\phi, \cos\theta) ~, 
  \\
  \hat{\bm{u}} &=(\sin\phi, -\cos\phi, 0) ~, \\
  \hat{\bm{v}} &=(\cos\theta\cos\phi, \cos\theta\sin\phi, -\sin\theta) ~.
\end{align}

Plugging Eq.~\eqref{eq:pwave} into Eqs.~\eqref{eq:z_time} and \eqref{eq:chi_time}, the redshift $z$ and the rotation of polarization angle $\chi$ can be written as 
\begin{align}
  z(t_{E}) &= \sum_{P = +, \times} \int_{-\infty}^{\infty} df \int d^2\hat{\bm{\Omega}} \,
  F^P(f,\hat{\bm{\Omega}})
  \notag \\
  &\quad \times \tilde{h}_P(f,\hat{\bm{\Omega}})\,e^{-2\pi i f t_{E} }~,
  \label{eq:z_time2}
  \\
  \chi(t_{E}) &= \sum_{P = +, \times} \int_{-\infty}^{\infty} df \int d^2\hat{\bm{\Omega}} \,
  R^P(f,\hat{\bm{\Omega}})
  \notag \\
  &\quad \times \tilde{h}_P(f,\hat{\bm{\Omega}})\,e^{-2\pi i f t_{E}} ~, 
  \label{eq:chi_times2}
\end{align}
where
\begin{align}\label{eq:respz}
  F^P(f,\hat{\bm{\Omega}}) &= 
  \frac{1}{2}\,
  \frac{\hat{n}^i\hat{n}^j e^P_{ij}(\hat{\bm{\Omega}})}{1+\hat{\bm{\Omega}}\cdot\hat{\bm n}}
  \big[ 1 - e^{2\pi i f L (1+\hat{\bm{\Omega}}\cdot\hat{\bm n})} \big]~,\\
  \label{eq:respchi}
  R^P(f,\hat{\bm{\Omega}}) &= 
  - \frac{1}{2}\,
  \frac{\big[ \hat{\Omega}^i e^P_{jk}(\hat{\bm{\Omega}}) - \hat{\Omega}^j e^P_{ik}(\hat{\bm{\Omega}}) \big] 
  \bar{\epsilon}^i \hat{m}^j \hat{n}^k 
  }
       {1+\hat{\bm{\Omega}}\cdot\hat{\bm n}} 
       \notag \\
    &\quad \times
    \big[ 1-e^{2\pi i f L (1+\hat{\bm{\Omega}}\cdot\hat{\bm n})} \big]~,
\end{align}
are the response functions for the redshift and the polarization rotation, respectively.
Note that the response functions satisfy the complementary relations,
\begin{align}\label{eq:respeq}
  R^{+}(f,\hat{\bm{\Omega}}) &= - F^{\times}(f,\hat{\bm{\Omega}}) ~,
  \quad R^{\times}(f,\hat{\bm{\Omega}}) = F^{+}(f,\hat{\bm{\Omega}}) ~.
\end{align}
For example, using the definitions of the polarization tensors \eqref{eq:polGW} and $\hat{\bm{m}} = \hat{\bm{n}} \times \bar{\bm{\epsilon}}$, the first relation can be derived as 
\begin{align}
	&(\hat{\Omega}^i e^+_{jk} -  \hat{\Omega}^j e^+_{ik}) 
    \bar{\epsilon}^i \hat{m}^j \hat{n}^k 
    \notag \\
    &= (\hat{\bm{\Omega}} \cdot \bar{\bm{\epsilon}})
    [ ( \hat{\bm{m}} \cdot \hat{\bm{u}} ) (\hat{\bm{n}} \cdot \hat{\bm{u}}) 
    - ( \hat{\bm{m}} \cdot \hat{\bm{v}} ) (\hat{\bm{n}} \cdot \hat{\bm{v}}) ] 
    \notag \\
    &\quad -
    (\hat{\bm{\Omega}} \cdot \hat{\bm{m}})
    [ ( \bar{\bm{\epsilon}} \cdot \hat{\bm{u}} ) (\hat{\bm{n}} \cdot \hat{\bm{u}}) 
    - ( \bar{\bm{\epsilon}} \cdot \hat{\bm{v}} ) (\hat{\bm{n}} \cdot \hat{\bm{v}}) ] 
    \notag \\
    &= - (\hat{\bm{n}} \cdot \hat{\bm{u}}) [ (\hat{\bm{\Omega}} \cdot \hat{\bm{m}}) ( \bar{\bm{\epsilon}} \cdot \hat{\bm{u}} ) - (\hat{\bm{\Omega}} \cdot \bar{\bm{\epsilon}}) ( \hat{\bm{m}} \cdot \hat{\bm{u}} ) ]
    \notag \\
    &\quad + (\hat{\bm{n}} \cdot \hat{\bm{v}})
    [ (\hat{\bm{\Omega}} \cdot \hat{\bm{m}}) ( \bar{\bm{\epsilon}} \cdot \hat{\bm{v}} ) - (\hat{\bm{\Omega}} \cdot \bar{\bm{\epsilon}}) ( \hat{\bm{m}} \cdot \hat{\bm{v}} ) ]
    \notag \\
    &= - (\hat{\bm{n}} \cdot \hat{\bm{u}}) [ (\hat{\bm{m}} \times \bar{\bm{\epsilon}}) \cdot (\hat{\bm{\Omega}} \times \hat{\bm{u}}) ]
    \notag \\
    &\quad + (\hat{\bm{n}} \cdot \hat{\bm{v}})
    [ (\hat{\bm{m}} \times \bar{\bm{\epsilon}}) \cdot (\hat{\bm{\Omega}} \times \hat{\bm{v}}) ]
    \notag \\
    &= 2 (\hat{\bm{n}} \cdot \hat{\bm{u}}) (\hat{\bm{n}} \cdot \hat{\bm{v}})
    \notag \\
    &=
    \hat{n}^i\hat{n}^j e^\times_{ij}~,
\end{align}
where we have used the identity $ (\bm{A} \cdot \bm{C}) (\bm{B} \cdot \bm{D}) - (\bm{B} \cdot \bm{C}) (\bm{A} \cdot \bm{D}) = (\bm{A} \times \bm{B}) \cdot (\bm{C} \times \bm{D})$ in the third equality.
The relations between $F^P$ and $R^P$ provide the geometric origin of the sensitivity to the parity-violating component, exploited in the cross-correlation analysis presented in Sec.~\ref{sec:3}.

\section{Correlation analysis with Pulsar timing and polarimetry}\label{sec:3}

\subsection{Cross-correlation created by SGWB}

In this section, we study how the measurements of the pulsar timing redshift and the rotation of polarization angle can be used to detect the SGWB. We model the SGWB as a stationary, isotropic, and Gaussian random field. Throughout this section, we adopt the polarization basis of GWs given in Eq.~\eqref{eq:polGW}. The statistical properties of the background are encoded in the two-point function of the Fourier amplitudes \cite{Seto:2008sr}, 
\begin{align}
    &\begin{pmatrix}
        \braket{\tilde{h}_+ (f,\hat{\bm{\Omega}}) \, \tilde{h}_{+}^*(f',\hat{\bm{\Omega}}')}
        &\braket{\tilde{h}_+ (f,\hat{\bm{\Omega}}) \, \tilde{h}_{\times}^*(f',\hat{\bm{\Omega}}')} \\
        \braket{\tilde{h}_\times (f,\hat{\bm{\Omega}}) \, \tilde{h}_{+}^*(f',\hat{\bm{\Omega}}')} 
        &\braket{\tilde{h}_\times (f,\hat{\bm{\Omega}}) \, \tilde{h}_{\times}^*(f',\hat{\bm{\Omega}}')}
    \end{pmatrix}
    \notag \\
	&= 
    \delta(f-f') \, \delta^2(\hat{\bm{\Omega}} - \hat{\bm{\Omega}}')
    \begin{pmatrix}
		 I(f) &  -i \,V(f)\\
		 i\, V(f) & I(f)
	\end{pmatrix} ,
    \label{twopoint}
\end{align}
where the angle brackets denote ensemble averages.
The $2\times2$ Hermitian matrix on the right-hand side defines the Stokes parameters of the SGWB. In particular, $I(f)$ represents the intensity, while $V(f)$ quantifies the circular polarization and thus encodes the parity violation of the SGWB. 
We omit the other Stokes parameters, $Q$ and $U$, which characterize the linear polarizations of the SGWB, since they do not contribute to the isotropic component~\cite{Seto:2008sr}. 
The intensity $I(f)$ is related to the GW energy density per unit logarithmic frequency, normalized by the critical density of the universe, $\Omega^I_{\rm GW}(f)$, through
\begin{align}
	\Omega^I_{\rm GW}(f) = \frac{32 \pi^3}{3H_0^2} f^3 I(f)~,
\end{align}
where $H_0$ is the Hubble constant. 
For the circular polarization parameter $V(f)$, we introduce the corresponding dimensionless parameter $\Omega^V_{\rm GW}(f)$ as
\begin{align}
	\Omega^V_{\rm GW}(f) = \frac{32 \pi^3}{3H_0^2} f^3 V(f)~,
\end{align}
in analogy with $\Omega^I_{\rm GW}(f)$, which describes the difference between left- and right-handed modes.

An effective method for detecting a stochastic background is correlation analysis~\cite{Flanagan:1993ix, Allen:1997ad}. 
In pulsar observations, we cross-correlate observables from distinct pulsars. 
The relevant observables in our case are the timing redshift $z$ and the rotation of  polarization angle $\chi$. 
We begin by considering the cross-correlation of the timing redshifts of pulsars $A$ and $B$.
From Eq.~\eqref{eq:z_time2}, the Fourier component of the redshift of pulsar $A$ is given by 
\begin{align}
    \tilde{z}_A(f) 
    = \sum_{P = +, \times} \int d^2\hat{\bm{\Omega}} \,
  F^P_A(f,\hat{\bm{\Omega}}) \,
   \tilde{h}_P(f,\hat{\bm{\Omega}}) \,,
\end{align}
where $F^P_A(f,\hat{\bm{\Omega}})$ is the response function of pulsar $A$'s redshift,
\begin{align}
    F^P_A(f,\hat{\bm{\Omega}}) &= 
  \frac{1}{2}\,
  \frac{\hat{n}_A^i\hat{n}_A^j e^P_{ij}(\hat{\bm{\Omega}})}{1+\hat{\bm{\Omega}}\cdot\hat{\bm n}_A}
  \big[ 1 - e^{2\pi i f L_A (1+\hat{\bm{\Omega}}\cdot\hat{\bm n}_A)} \big] ~,
  \label{eq:FAP}
\end{align}
with $L_A$ denoting the distance to pulsar $A$, and $\hat{\bm{n}}_A$ the unit vector pointing from the Earth to pulsar $A$.
From Eq.~\eqref{twopoint}, the cross-correlation of the redshifts of pulsar $A$ and $B$ is
\begin{align}
	\braket{\tilde{z}_A(f) \, \tilde{z}_B^*(f')} &= \delta(f-f') \big[
    \Gamma^I_{AB}(f) \, I(f) + \Gamma^V_{AB}(f) \, V(f) \big]~.
\end{align}
Here, we introduced the overlap reduction functions (ORFs),
\begin{align}
	\Gamma^I_{AB}(f) &\equiv 
    \int d^2 \hat{\bm{\Omega}} \, \big[ 
    F^+_A(f,\hat{\bm{\Omega}}) \,F^{+*}_B(f,\hat{\bm{\Omega}}) 
    \notag \\
    &\qquad + F^\times_A(f,\hat{\bm{\Omega}}) \, F^{\times *}_B(f,\hat{\bm{\Omega}})
    \big]~,\\
	\Gamma^V_{AB}(f) &\equiv - i \int d^2 \hat{\bm{\Omega}} \, \big[ F^+_A(f,\hat{\bm{\Omega}}) \, F^{\times *}_B(f,\hat{\bm{\Omega}}) 
    \notag \\
    &\qquad 
    - F^\times_A(f,\hat{\bm{\Omega}}) \, F^{+ *}_B(f,\hat{\bm{\Omega}}) \big]~.
\end{align}
In pulsar systems, where $f \sim 10^{-9}\,\text{Hz}$ and $L \sim 10^{2}$--$10^3 \, \text{pc}$, the factor $2\pi f L $ is typically large. 
In this long-distance regime, the pulsar term, represented by the exponential factor in Eq.~\eqref{eq:FAP}, oscillates rapidly and therefore averages out in the two-point correlation. Therefore, both the parity-even ORF $\Gamma^I_{AB}$ and the parity-odd ORF $\Gamma^V_{AB}$ become independent of the frequency $f$, with $\Gamma^I_{AB}$ reducing to the Hellings--Downs curve~\cite{Hellings:1983fr} and $\Gamma^V_{AB}$ vanishing~\cite{Kato:2015bye}.
\begin{align}\label{eq:ORFz}
	\Gamma^I_{AB} &= \frac{8\pi}{3}\Gamma_{\rm HD}(\xi_{AB})~, & \Gamma^V_{AB} &= 0~.
\end{align}
Here $\Gamma_{\rm HD}(\xi_{AB})$ is a function of the separation angle $\xi_{AB}$ between pulsars $A$ and $B$, given by
\begin{align}
	\Gamma_{\rm HD}(\xi) &= \frac{1}{2} - \frac{1}{4}\left(\frac{1-\cos\xi}{2}\right)\cr
    &\qquad + \frac{3}{2}\left(\frac{1-\cos\xi}{2}\right) \ln \left(\frac{1-\cos\xi}{2}\right) ~.
\end{align}
Therefore, we obtain 
\begin{align}
	\braket{\tilde{z}_A(f) \, \tilde{z}_B^*(f')} &= \delta(f-f') \, \frac{8 \pi}{3} \Gamma_{\text{HD}}(\xi_{AB})
    \, I(f) ~.
\end{align}

The vanishing of $\Gamma^{V}_{AB}$ reflects the parity symmetry of the configuration: the Earth and the two pulsars always lie in the same plane, making the system invariant under a parity transformation.
Since the circular polarization $V$ is parity-odd, its response necessarily vanishes.

From Eq.~\eqref{eq:chi_times2}, the Fourier component of the rotation of polarization angle of pulsar $A$ is
\begin{align}
    \tilde{\chi}_A(f) &= \sum_{P = +, \times} \int d^2\hat{\bm{\Omega}} \,
  R^P_A(f,\hat{\bm{\Omega}}) \, 
  \tilde{h}_P(f,\hat{\bm{\Omega}})~.
\end{align}
Here, $R^P_A(f,\hat{\bm{\Omega}})$ is the response function of pulsar $A$'s polarization rotation, related to $F^P_A(f,\hat{\bm{\Omega}})$ through $R_A^{+} = - F_A^{\times}$, $R_A^{\times} = F_A^{+}$ as in Eq.~\eqref{eq:respeq}.
Using these relations, the other cross-correlations can be obtained as
\begin{align}	\braket{\tilde{\chi}_A(f) \, \tilde{\chi}_B^*(f')} &= \delta(f-f') \, \frac{8\pi}{3} \, \Gamma_{\rm HD}(\xi_{AB}) \, I(f)~, \\ \braket{\tilde{z}_A(f) \, \tilde{\chi}_B^*(f')} &= -i\,\delta(f-f') \,
    \frac{8\pi}{3} \, \Gamma_{\rm HD}(\xi_{AB}) \, V(f)~.
    \label{eq:z-chi-cor}
\end{align}

Notably, the $z$-$\chi$ cross-correlation is purely imaginary and depends only on $V(f)$. This parity-odd estimator enables a clean separation of the circular polarization $V$ from the intensity $I$. 
Moreover, the angular correlation pattern consistently reduces to the Hellings--Downs curve. As discussed in Ref.~\cite{Kehagias:2024plp}, the Hellings--Downs curve arises from Lorentz symmetry, which reflects the conformal properties of the two-point correlation of scalar quantities on the celestial sphere.

\subsection{Sensitivity forecast}

So far, we have only considered contributions from the SGWB. 
In practice, each observable consists of a signal induced by the GWs and an additional stochastic noise contribution. 
For each pulsar $A$, we work in the Fourier domain and define the measured quantities $\tilde{d}_A^{(z)}(f)$ for the timing redshift and $\tilde{d}_A^{(\chi)}(f)$ for polarimetry as
\begin{align}
	\tilde{d}^{(z)}_A(f) &= \tilde{z}_A(f) + \tilde{n}^{(z)}_A(f)~, \cr
	\tilde{d}^{(\chi)}_A(f) &= \tilde{\chi}_A(f) + \tilde{n}^{(\chi)}_A(f)~,
\end{align}
where $\tilde{n}^{(z)}_A(f)$ and $\tilde{n}^{(\chi)}_A(f)$ represent the noise contributions in the timing redshift and polarimetry measurements, respectively.
The signal components $\tilde{z}_A(f)$ and $\tilde{\chi}_A(f)$ obey the ensemble relations derived in the previous subsection, while the noise terms $\tilde{n}_A^{(z)}(f)$ and $\tilde{n}_A^{(\chi)}(f)$ account for instrumental and propagation effects that are not modeled explicitly.
We assume that the noise is stationary, Gaussian, and uncorrelated between different pulsars. Under this assumption, it can be characterized by one-sided power spectral densities (PSDs) for each measurement channel,
\begin{align}
	\braket{\tilde{n}^{(z)}_A(f) \,\tilde{n}^{(z)*}_B(f')} &= \frac{1}{2} \,\delta(f-f') \, \delta_{AB} \, S^{(z)}_A(f)~, \\
    \braket{\tilde{n}^{(\chi)}_A(f) \, \tilde{n}^{(\chi)*}_B(f')} &= \frac{1}{2} \, \delta(f-f') \, \delta_{AB} \, S^{(\chi)}_A(f)~.
\end{align}
We further assume that cross-channel couplings are negligible, so that the timing and polarimetry noises are uncorrelated:
\begin{align}
    \braket{\tilde{n}^{(z)}_A(f) \,\tilde{n}^{(\chi)*}_B(f')} &= 0 \,.
\end{align}

For simplicity, we consider the case of white noise for both the pulsar timing residuals (i.e., the time integral of the redshift) and the pulsar polarimetry measurements.
We assume that all pulsars are observed with the same cadence $\delta t$, and that the timing residuals and polarimetry measurements have root-mean-square (rms) uncertainties of $\sigma_{t}$ and $\sigma_{p}$, respectively.
The corresponding one-sided PSDs are given by
\begin{align}
	S_A^{(z)}(f) &=  2 (2 \pi f \sigma_t)^2 \, \delta t~, & S_A^{(\chi)}(f) &= 2 \sigma_p^2 \, \delta t~.
\end{align}
Intrinsic spin noise or contributions from the interstellar medium can be incorporated into $S_A^{(z)}(f)$ or $S_A^{(\chi)}(f)$ with their measured or modeled frequency dependence; however, we neglect them here~\footnote{The effect of the rotation induced by the interstellar medium and the Earth's ionosphere has been considered in axion searches with pulsar polarization arrays~\cite{PPTA:2024mgh}. A similar estimate can in principle be performed in our case, but we leave this for future work.}.

Looking ahead, timing measurements are expected to reach $\delta t \sim 2\,\text{weeks}$ and $\sigma_t \sim 30 \,{\rm ns}$ in the SKA era~\cite{Janssen:2014dka, Moore:2014lga}. 
For polarimetry, recent observations already achieve $\sigma_p \sim 0.5 \, \text{deg}$ per epoch~\cite{2025AJ....170..116C}.
For forecasts, we may utilize an optimistic uncertainty $\sigma_p \sim 0.1\, \text{deg}$. With these values, the noise level of pulsar polarimetry is approximately $10^{23}$ times larger than that of pulsar timing at $f = 1/{\rm yr}$.

To target the circular polarization of the SGWB, $V$, we construct a cross-correlation statistic between the timing and polarimetry measurements of distinct pulsars.
Each frequency bin is weighted by the inverse noise PSDs and by the overlap reduction function, which encodes the angular dependence of the pulsar pairs.
Following the procedure for computing the optimal cross-correlation statistic in PTAs~\cite{Anholm:2008wy, Siemens:2013zla}, the signal-to-noise ratio (SNR) for the Stokes parameter $V$ can be estimated as
\begin{align}\label{eq:SNR}
	{\rm SNR}^2 &\sim \left(\frac{H_0^2}{4 \pi^2 }\right)^2 
    \frac{N_p^2 T_{\rm obs}}{24}\int^{1/2\delta t}_{1/T_{\rm obs}} \frac{df}{f^6}\frac{\left(\Omega^{V}_{\rm GW}(f)\right)^2}{S^{(z)}(f)S^{(\chi)}(f)}~,
\end{align}
where $T_{\text{obs}}$ represents the observation time span, and $N_p$ is the number of pulsars.
This expression assumes that the noise power spectra are identical for all pulsars and that the number of pulsars is large, $N_p \gg 1$. 
In this regime, the summation over pulsar pairs $\sum_{A<B}\Gamma_{\rm HD}(\xi_{AB})^2$ can be approximated by an integral, which leads to the factor $1/24$ in Eq.~\eqref{eq:SNR}. 

If the specturm is flat, $\Omega^V_{\rm GW}(f)$ is independent of $f$, the SNR roughly reads
\begin{align}
	{\rm SNR} &\sim  5 \left(\frac{\Omega^V_{\rm GW}}{0.012}\right)\left(\frac{T_{\rm obs}}{20 \, {\rm yr}}\right)^4 \left(\frac{N_p}{200}\right) \left(\frac{\sigma_t}{30 \, {\rm ns}}\right)^{-1} \cr
    &\qquad \times \left(\frac{\sigma_p}{0.1 \, {\rm deg}}\right)^{-1}\left(\frac{\delta t}{2 \, {\rm week}}\right)^{-1},
\end{align}
where we have used $H_0 = 70 \, \text{km/s/Mpc}$. 
Thus, the PTA-PPA cross-correlation would place a limit $\Omega_{\rm GW}^V \lesssim 0.012$ on the Stokes parameter $V$ of the SGWB, which might be achievable in the SKA era.

\section{Conclusion}
\label{sec:4}
We have investigated the rotation of electromagnetic wave polarization during propagation through GWs and explored the joint use of pulsar polarimetry and timing measurements to detect circular polarization of the SGWB, which would indicate parity violation.
Starting from geometric optics in a GW background, we derived the perturbed photon geodesic and the transport equation for the polarization vector, obtaining expressions for the GW-induced rotation of polarization angle as well as for the timing redshift.
The rotation of polarization angle is a pseudo-scalar under parity transformation, and its antenna pattern is related to that of the timing response through the exchange of the GW's $+$ and $\times$ polarizations.

Assuming a stationary, isotropic, and Gaussian SGWB, we computed the cross-correlations constructed from the pulsar timing redshift $z$ and the rotation of polarization angle $\chi$.
We found that the $z$-$z$ and $\chi$-$\chi$ correlations depend solely on the SGWB intensity $I$, whereas the mixed $z$-$\chi$ correlation isolates the circular polarization component $V$. These correlations share the same Hellings--Downs angular pattern.
Consequently, combining pulsar timing and polarimetry measurements allows separate inference of the SGWB intensity $I$ and circular polarization $V$, without requiring any assumptions about their relative amplitudes.

The detection significance of the $z$-$\chi$ correlation is determined by the timing and polarimetry noise power spectral densities and the effective bandwidth in Eq.~\eqref{eq:SNR}, particularly the lower cutoff frequency.
Adopting optimistic parameters ($N_p\sim 200, T_{\rm obs}\sim 20~{\rm yr}, \delta t \sim 2~{\rm week}, \sigma_t\sim 30~{\rm ns}$ and $\sigma_p\sim 0.1~{\rm deg}$) yields a constraint of $\Omega_{\rm GW}^V \lesssim 0.012$.
Although this sensitivity appears lower than that of current PTA measurements for $\Omega_{\text{GW}}^I$, our method provides a valuable null test with a sensitivity comparable to the current constraint from astrometry survey~\cite{Jaraba:2025hay}. 
These estimates should be regarded as order-of-magnitude. Although the forecasts lack the precision required for a clear detection, our study lays the groundwork for investigating the parity-violation signal of the SGWB in the PPA system, and the methodology can be extended to other stellar objects with well-characterized polarization and timing models. 

\begin{acknowledgments}
We thank Yifan Chen and Claudia de Rham for useful discussions. This work was initiated at the Aspen Center for Physics, which is supported by the National Science Foundation under Grant No. PHY-2210452 and by a grant from the Simons Foundation (No. 1161654, Troyer). 
Q.L. is supported by World Premier International Research Center Initiative (WPI Initiative), MEXT, Japan.
K.N. was supported by JSPS KAKENHI Grant Numbers JP24KJ0117 and JP25K17389. 
H. O. was supported by JSPS KAKENHI Grant Numbers JP23H00110, 25K17388, and Yamada Science Foundation. 
\end{acknowledgments}

\appendix

\section{Geometric optics in curved spacetime}
\label{app:geometric-optics}

In this Appendix, we briefly review geometric optics in curved spacetime and derive the equations of light propagation, Eqs.~\eqref{eq:geodlight} and \eqref{eq:polprop}.
For a detailed discussion, see, e.g., Sec.~22.5 of Ref.~\cite{Misner:1973prb}.

We start from the Maxwell equation in vacuum, $\nabla_\nu F^{\nu\mu} = 0$.
Here, $\nabla_\mu$ denotes the covariant derivative associated with the spacetime metric $g_{\mu\nu}$, and $F_{\mu\nu} \equiv \nabla_\mu A_\nu - \nabla_\nu A_\mu$ is the electromagnetic field strength tensor with $A_\mu$ being the gauge field.
For the gauge field, we impose the Lorenz gauge condition,
\begin{align}
    \nabla_\mu A^\mu &= 0 ~. 
    \label{eq:lorenz}
\end{align}
In this gauge, the Maxwell equation $\nabla_\nu F^{\nu\mu} = 0$ can be rewritten as 
\begin{align}
    \nabla^\nu \nabla_\nu A^\mu - {R^\mu}_\nu A^\nu &= 0 ~,
    \label{eq:geoopt}
\end{align}
where $R_{\mu\nu}$ denotes the Ricci tensor.

Geometric optics is valid when the wavelength of light, $\lambda$, is much shorter than a characteristic length scale $L$, which represents the curvature radius of the background spacetime or a typical scale over which the amplitude, polarization, and wavelength of the light vary.
It is convenient to formally express the gauge field as
\begin{align}
    A^\mu (x) = [ a^\mu (x) + s\, b^\mu (x) + \cdots  ] \, e^{i\theta(x) / s} ~. 
    \label{eq:amuexp}
\end{align}
Here, $\theta(x)$ is a real function representing the phase of the electromagnetic wave, while $a^\mu(x)$, $b^\mu(x)$, and so on are, in general, complex vector functions.
We have also introduced an auxiliary parameter $s$, which is ultimately set to unity, to indicate that terms proportional to $s^n$ scale as $(\lambda / L)^n$ in the short-wavelength limit.

The wave vector of the electromagnetic wave, $k^\mu$, is defined as the gradient of the phase: $k^\mu \equiv \nabla^\mu \theta$.
It is also conventional to decompose the vector function $a^\mu$ as $a^\mu = a \, \epsilon^\mu$, where $a \equiv \sqrt{a^\mu a^*_\mu}$ denotes the scalar amplitude and $\epsilon^\mu$ is the polarization vector.
(The asterisk indicates complex conjugation.)
By definition, the polarization vector is normalized as $\epsilon^\mu \epsilon^*_\mu = 1$.

Substituting the expansion \eqref{eq:amuexp} into the Lorenz gauge condition \eqref{eq:lorenz}, we obtain, at leading order, $k_\mu a^\mu = 0$, or, equivalently,
\begin{align}
k_\mu \epsilon^\mu &= 0~,
\end{align}
which is the second equation in Eq.~\eqref{eq:polprop}.
On the other hand, substituting Eq.~\eqref{eq:amuexp} into the Maxwell equation \eqref{eq:geoopt} yields the null condition $k_\mu k^\mu = 0$ at leading order.
Taking the derivative of this equation and noting that $\nabla_\mu k_\nu = \nabla_\nu k_\mu$ follows from $k_\mu = \nabla_\mu \theta$, we find
\begin{align}
    k^\nu \nabla_\nu k^\mu &= 0 ~,
\end{align}
which is the geodesic equation \eqref{eq:geodlight}.

At the next order in $s$, from Eq.~\eqref{eq:geoopt} we obtain
\begin{align}
    k^\nu \nabla_\nu a^\mu = -\frac{1}{2} a^\mu (\nabla_\nu k^\nu)~. 
    \label{eq:geoopt21}
\end{align}

From the definition $a^\mu = a \, \epsilon^\mu$ with $a = \sqrt{a^\mu a^*_\mu}$, the above equation can be recast into the evolution equations for the scalar amplitude and the polarization vector as
\begin{align}
k^\nu \nabla_\nu a &= -\frac{1}{2} a\, (\nabla_\nu k^\nu) ~,
\label{eq:geoopt22}\\
k^\nu \nabla_\nu \epsilon^\mu &= 0~.
\label{eq:geoopt23}
\end{align}
Equation~\eqref{eq:geoopt23} is the first of Eq.~\eqref{eq:polprop}, indicating that the polarization vector is parallel transported along the light ray.

\bibliography{ref}

\end{document}